\documentclass[aps,amsmath,amssymb,nofootinbib]{revtex4}

\usepackage{graphicx}
\usepackage{dcolumn}
\usepackage{bm}

\def\eff{{\rm eff}}
\def\emm{{\rm EM}}
\def\EH{{\rm E-H}}
\def\intt{{\rm int}}
\def\t{\tilde}

\begin{document}

\title{Electromagnetic Force on a Brane} 

\author{Li-Xin Li}
\email{lxl@pku.edu.cn}
\affiliation{Kavli Institute for Astronomy and Astrophysics, Peking University, Beijing 100871, P. R. China}

\date{\today}

\begin{abstract}
A fundamental assumption in the theory of brane world is that all matter and radiation are confined on the four-dimensional brane and only gravitons can propagate in the five-dimensional bulk spacetime. The brane world theory did not provide an explanation for the existence of electromagnetic fields and the origin of the electromagnetic field equation. In this paper, we propose a model for explaining the existence of electromagnetic fields on a brane and deriving the electromagnetic field equation. Similar to the case in Kaluza-Klein theory, we find that electromagnetic fields and the electromagnetic field equation can be derived from the five-dimensional Einstein field equation. However, the derived electromagnetic field equation differs from the Maxwell equation by containing a term with the electromagnetic potential vector coupled to the spacetime curvature tensor. So it can be considered as generalization of the Maxwell equation in a curved spacetime. The gravitational field equation on the brane is also derived with the stress-energy tensor for electromagnetic fields explicitly included and the Weyl tensor term explicitly expressed with matter fields and their derivatives in the direction of the extra-dimension. The model proposed in the paper can be regarded as unification of electromagnetic and gravitational interactions in the framework of brane world theory.

\vspace{0.4cm}
\end{abstract}

\maketitle

\section{Introduction}

The theory of brane world was proposed to address the hierarchy problem in theoretical physics \cite{ran99a,ran99b}. In the theory of brane world, the four-dimensional spacetime in which we live is assumed to be a hypersurface (or, a brane) embedded in a five-dimensional bulk spacetime. The gravitational field equation on the four-dimensional brane is derived from the Einstein field equation in the five-dimensional bulk space by the approach of projection \cite{shi00}. Standard model particles, including electromagnetic fields, strong and weak particles, are assumed to be confined on the four-dimensional brane. The assumption is motivated by D-branes in string theory, on which open strings representing the non-gravitational sector can end \cite{hor96,luk99,luk99b,luk00}. For a recent and comprehensive review on the theory of brane world and its application in physics and cosmology, please refer to \cite{maa10}.

Kaluza-Klein (KK) theory also attempts to interpret the physics in a four-dimensional spacetime as arising from the gravity in a five-dimensional bulk spacetime \cite{kal21,kle26a,kle26b,bai87,ove97}. In the KK theory, both the Maxwell equation and the four-dimensional Einstein field equation are derived from the five-dimensional Einstein field equation. In the KK theory the extra-dimension is assumed  to be compactified to a very small scale so that the extra-dimension cannot be seen in normal physical experiments and hence gravity appears four-dimensional. However, in the brane theory the extra-dimension can be noncompact. It is the curvature of the bulk space that keeps gravity to be four-dimensional on scales larger than the curvature radius of the bulk space \cite{ran99b}.

Although both attempt to interpret the four-dimensional physics as arising from the five-dimensional physics, KK and brane world theories are distinctly different in physics: they are defined on two different hypersurfaces in a five-dimensional spacetime and are not related by diffeomorphisms, as explained in detail in \cite{li15a}. As can be seen from the above description, electromagnetism has different origins in the two theories. In the KK theory, the Maxwell equation is derived from the five-dimensional Einstein field equation, hence electromagnetism and gravity have the same origin. In the brane world theory, electromagnetism is assumed to arise from open strings ending on D-branes hence has no relation to gravity arising from closed strings. In addition, as mentioned above, in the KK theory the extra spatial dimension must be compactified, but in the brane theory the extra spatial dimension can be noncompact.

It would be interesting to adapt the idea of deriving the Maxwell equation from the five-dimensional Einstein field equation in the KK theory to the brane world theory. However, this is not an easy task, since the 4+1 decomposition of the five-dimensional metric tensor adopted in the KK theory is different from that used for obtaining a metric tensor on a brane hypersurface, as explained in \cite{li15a}. In this paper, we propose a brane world model in which electromagnetic fields and the electromagnetic field equation on the brane are derived from the five-dimensional Einstein field equation. We will see that, this can be realized only if an appropriate boundary condition on the brane is adopted, and the derived electromagnetic field equation differs from the Maxwell equation by a curvature-coupled term. The boundary condition differs from the $Z_2$-symmetry boundary condition used in the standard brane world theory, as will be explained in the paper.

\section{4+1 Decomposition of the Five-dimensional Einstein Field Equation}

Assuming a five-dimensional spacetime in which gravity is described by the Einstein field equation
\begin{eqnarray}
        \t{R}_{ab}-\frac{1}{2}\t{R}\t{g}_{ab}=\t{\kappa}\t{T}_{ab} \;, \label{ein_eq5}
\end{eqnarray}
where $\t{g}_{ab}$ is the metric tensor of the spacetime, $\t{R}_{ab}$ is the Ricci curvature tensor, $\t{R}\equiv \t{g}^{ab}\t{R}_{ab}$ is the Ricci scalar, $\t{T}_{ab}$ is the stress-energy tensor of matter, and $\t{\kappa}$ is gravitational coupling constant. The five-dimensional spacetime is sliced by a set of timelike hypersurfaces, and one of the hypersurfaces is just the four-dimensional spacetime in which we live. The hypersurface has a unit normal $n^a$, a metric tensor $g_{ab}\equiv \t{g}_{ab}-n_an_b$, and an extrinsic curvature tensor $K_{ab}\equiv\t{\pounds}_ng_{ab}$ where $\t{\pounds}_n$ denotes the Lie derivative with respect to $n^a$. 

It is well known that the Einstein field equation (\ref{ein_eq5}) is equivalent to the following three equations expressed in terms of geometric quantities on the hypersurface \cite{mis73,wal84,shi00,li15a}: 
\begin{eqnarray}
        R +K_{ab}K^{ab}-K^2 = -2\t{\kappa}\t{T}_{ab}n^an^b \;, \label{scalar_eq}
\end{eqnarray}
\begin{eqnarray}
        \nabla_aK^{ab}-\nabla^bK = \t{\kappa}g^{ab}\t{T}_{ac}n^c \;, \label{vector_eq}
\end{eqnarray}
and
\begin{eqnarray}
	G_{ab} = \t{\kappa}g_a^{\;\;c}g_b^{\;\;d}\t{T}_{cd} +g_{ab}{}{}^{cd}\t{\pounds}_nK_{cd} -\left(2K_a^{\;\;c}K_{cb}-KK_{ab}\right)+\frac{1}{2}\left(3K_{cd}K^{cd}-K^2\right)g_{ab} \;. \label{tensor_eq}
\end{eqnarray}
Here, $g_{abcd}\equiv g_{ac}g_{bd}-g_{ab}g_{cd}$, $\nabla_a$ is the derivative operator associated with the metric $g_{ab}$, $R_{ab}$ is the Ricci curvature tensor of the brane hypersurface, $R\equiv R_a{}^a$ is the Ricci scalar, $G_{ab}\equiv R_{ab}-(1/2)Rg_{ab}$ is the Einstein tensor, and $K\equiv K_a{}^a$. 

The scalar equation (\ref{scalar_eq}) is obtained by contraction of equation (\ref{ein_eq5}) with $n^an^b$. The vector equation (\ref{vector_eq}) is obtained by contraction of equation (\ref{ein_eq5}) with $n^b$, then projection onto the hypersurface by the projection operator $g_{ab}$. It is also called the Gauss-Codacci relation \cite{mis73,wal84}. The tensor equation (\ref{tensor_eq}) is obtained by full projection of the equation (\ref{ein_eq5}) onto the hypersurface. We use normal letters to denote quantities on the brane hypersurface, and tilded letters to denote quantities defined in the bulk spacetime (except the normal vector $n^a$). The index of a tensor in the bulk spacetime is lowered (raised) by $\t{g}_{ab}$ ($\t{g}^{ab}$). The index of a tensor on the brane can be lowered (raised) by both $g_{ab}$ ($g^{ab}$) and $\t{g}_{ab}$ ($\t{g}^{ab}$), with the same result. 

In the brane world theory with the $Z_2$-symmetry boundary condition, the tensor field equation (\ref{tensor_eq}) is interpreted as the Einstein field equation on the bane \cite{shi00}. Equation (\ref{tensor_eq}) agrees with the eq.~8 of \cite{shi00}, if the traceless tensor $E_{ab}$ in the eq.~8 in \cite{shi00}, which is defined by the Weyl tensor in the bulk space, is expressed in terms of the brane extrinsic curvature $K_{ab}$ and its derivative in the direction orthogonal to the brane. Note that, in derivation of equation (\ref{tensor_eq}), following \cite{shi00} we have assumed that $n^a$ is tangent to a geodesic so that the acceleration vector $a^b\equiv n^a\t{\nabla}_an^b=0$.

\section{Electromagnetic Field Equations on a Brane}

Let us consider a discontinuous hypersurface (a brane) in a five-dimensional spacetime, which contains a surface stress-energy tensor as assumed in the theory of brane world. To have a well-defined four-dimensional geometry on the brane, the induced metric $g_{ab}$ must be continuous across it, hence the derivative operator $\nabla_a$ and the Riemann curvature defined by it. Then, by definition, the extrinsic curvature $K_{ab}$ does not contain a Dirac $\delta$-function, although it can be discontinuous across the brane. In fact, by the five-dimensional Einstein field equation (\ref{ein_eq5}), $K_{ab}$ must be discontinuous across the brane.

The bulk stress-energy tensor $\t{T}_{ab}$ must contain a $\delta$-function at the position of the brane. So, we can write
\begin{eqnarray}
	\t{T}_{ab}=-\frac{\t{\Lambda}}{\t{\kappa}}\t{g}_{ab}+\t{\cal T}_{ab}+\t{S}_{ab}\delta(n) \;, \label{tT_tSab}
\end{eqnarray}
where $\t{\Lambda}$ is the cosmological constant in the bulk space, $\t{\cal T}_{ab}$ and $\t{S}_{ab}$ are regular tensors (i.e., they contain no $\delta$-function). Here we have written $n^a=(\partial/\partial n)^a$ and use $n=0$ to denote the position of the brane. 

Let $[Q]$ denote the difference in the value of any quantity $Q$ on the two sides of the brane, i.e., $[Q]\equiv Q^+-Q^-$, $Q^+=Q(n=0^+)$, and $Q^-=Q(n=0^-)$. Integration of equation (\ref{tensor_eq}) across the brane hypersurface leads to the Israel junction condition \cite{isr66,mis73}
\begin{eqnarray}
	 [K_{ab}]=-\t{\kappa}\left(S_{ab}-\frac{1}{3}Sg_{ab}\right) \;, \label{join1}
\end{eqnarray}
where $S_{ab}\equiv g_a^{\;\;c}g_b^{\;\;d}\t{S}_{cd}$ and $S=S_a{}^a$. Similarly, integration of equations (\ref{scalar_eq}) and (\ref{vector_eq}) across the brane leads to
\begin{eqnarray}
	 n^cn^d\t{S}_{cd}=0 \hspace{0.5cm} \mbox{and} \hspace{1cm} g_b^{\;\;d}n^c\t{S}_{cd}=0 \;,
\end{eqnarray}
which simply tells that the momentum flow represented by $\t{S}_{ab}$ is entirely in the brane \cite{mis73}.

To introduce electromagnetic fields on the brane, in the neighborhood of the brane hypersurface we choose a general coordinate system $\{x^0,x^1,x^2,x^3,x^4\equiv w\}$ so that the brane is located at $w=0$. The coordinate vector $w^a=(\partial/\partial w)^a$ can be decomposed as $w^a=Nn^a+N^a$, where $N$ is the lapse function, and $N^a=g^a{}_bw^b$ is the shift vector \cite{wal84}. It can be verified that the acceleration vector $a^a=-\nabla^a\ln N$. Hence, the geodesic condition $a^a=0$ is identical to the condition $\nabla_aN=0$, i.e., $N$ can only be a function of $w$. Define
\begin{eqnarray}
	w^\prime=2\int Ndw \;, \hspace{0.5cm} 
        A^a=(2N)^{-1}N^a \;, \hspace{0.5cm}
        H_{ab}=\nabla_aA_b+\nabla_bA_a=H_{ba} \;,
\end{eqnarray}
the extrinsic curvature $K_{ab}$ can be expressed as
\begin{eqnarray}
	K_{ab} = \dot{g}_{ab}-H_{ab} \;, \label{Kab_H}
\end{eqnarray}
where $\dot{g}_{ab}\equiv\partial g_{ab}/\partial w^\prime\equiv g_a^{\;\;c}g_b^{\;\;d}\t{\pounds}_{w^\prime}g_{cd}$.

Substituting equation (\ref{Kab_H}) into the vector equation (\ref{vector_eq}), we get
\begin{eqnarray}
        \nabla_aF^{ab}+2R^b{}_aA^a=-4\pi J^b \;, \label{meq1}
\end{eqnarray} 
where $F_{ab}\equiv\nabla_aA_b-\nabla_bA_a$, 
\begin{eqnarray}
        J^a\equiv\frac{1}{4\pi}\left(\nabla_b\Phi^{ab}+\t{\kappa}g^{ab}\t{T}_{bc}n^c\right) \;, \label{Ja}
\end{eqnarray}
and
\begin{eqnarray}
	\Phi_{ab}\equiv -g_{ab}{}{}^{cd}\dot{g}_{cd}=\Phi_{ba} \;. \label{Phi_ab}
\end{eqnarray}
Equation (\ref{meq1}) differs from the Maxwell equation only by a curvature-coupled term $2R^b{}_aA^a$, if $A^a$ is interpreted as the electromagnetic potential vector, and $J^a$ interpreted as the electric current density vector. When $R_{ab}=0$, it is exactly the Maxwell equation. Therefore, we can interpret equation (\ref{meq1}) as generalization of the Maxwell equation in the brane world theory.

\section{Einstein Field Equations on a Brane}

Substituting equation (\ref{Kab_H}) in to the tensor equation (\ref{tensor_eq}), we get a four-dimensional Einstein field equation on the brane:
\begin{eqnarray}
	G_{ab}=\kappa T_{ab} \;, \label{ein_eq40}
\end{eqnarray}
with the stress-energy tensor
\begin{eqnarray}
	T_{ab} = T_{\emm,ab}+T_{m,ab}+T_{\intt,ab}+\frac{\t{\kappa}}{\kappa}g_a{}^cg_b{}^d\t{T}_{cd} \;, \label{Tab}
\end{eqnarray}
where
\begin{eqnarray}
	\kappa T_{\emm,ab} = 2\Psi_{ac}\Psi_b^{\;\;c}-\frac{2}{3}\Psi\Psi_{ab} -\frac{1}{2}\left(\Psi_{cd}\Psi^{cd}-\frac{1}{3}\Psi^2\right)g_{ab} -2\nabla^c\left(2A_{(a}\Psi_{b)c}-A_c\Psi_{ab}\right) \;, \label{Tab_em_M}
\end{eqnarray}
\begin{eqnarray}
	\kappa T_{m,ab} = -2\Phi_{ac}\Phi_b^{\;\;c} +\frac{1}{3}\Phi\Phi_{ab}-\frac{1}{2}\left(\Phi_{cd}\Phi^{cd}-\frac{1}{3}\Phi^2\right)g_{ab} -2\dot{\Phi}_{ab} \;, \label{Tab_m_M}
\end{eqnarray}
and
\begin{eqnarray}
	\kappa T_{\intt,ab} = \frac{1}{3}\Phi\Psi_{ab}-\frac{2}{3}\Psi\Phi_{ab}-\left(\Phi_{cd}\Psi^{cd}-\frac{1}{3}\Phi\Psi\right)g_{ab} -2\nabla^c\left(2A_{(a}\Phi_{b)c}-A_c\Phi_{ab}\right)-2\dot{\Psi}_{ab} \;. \label{Tab_int_M}
\end{eqnarray}
Here, $\kappa$ is the gravitational coupling constant in the four-dimensional spacetime,
\begin{eqnarray}
	\Psi_{ab}\equiv H_{ab}-Hg_{ab}=\Psi_{ba} \;, \label{Psi_ab}
\end{eqnarray}
$H\equiv H_c{}^c$, $\Psi\equiv \Psi_c{}^c$, and $\Phi\equiv \Phi_c{}^c$. The braces in tensor indexes denote symmetrization of indexes.

According to \cite{li15a}, $T_{\emm,ab}$ is interpreted as the stress-energy tensor of electromagnetic fields, $T_{m,ab}$ interpreted as the stress-energy tensor of the matter field associated with $\dot{g}_{ab}$, and $T_{\intt,ab}$ interpreted as the stress-energy tensor arising from the interaction between the electromagnetic field and the matter field. The $T_{\emm,ab}$ is related to the ordinary electromagnetic stress-energy tensor
\begin{eqnarray}
	\!\!~^{(0)}T_{\emm,ab}=\frac{2}{\kappa}\left(F_{ac}F_b{}^c-\frac{1}{4}g_{ab}F_{cd}F^{cd}\right) \label{T_em_0}
\end{eqnarray}
by
\begin{eqnarray}
	T_{\emm,ab} =\, ^{(0)}\!T_{\emm,ab} + ^{(1)}\!T_{\emm,ab} \;,
\end{eqnarray}
where
\begin{eqnarray}
        ^{(1)}T_{\emm,ab} = -\frac{2}{\kappa}\left\{\nabla^c\nabla_c(A_aA_b)-2\nabla^c\nabla_{(a}(A_{b)}A_c)+4A^cR_{c(a}A_{b)}+g_{ab}\left[\nabla_c\nabla_d(A^cA^d)-R_{cd}A^cA^d\right]\right\}, \hspace{0.3cm} \label{T_em_1}
\end{eqnarray}
which arises from the curvature-coupled term in the field equation (\ref{meq1}).

Unlike $\!\!~^{(0)}T_{\emm,ab}$, the $\!\!~^{(1)}T_{\emm,ab}$ does not interact with electric charge and current, since
\begin{eqnarray}
        \nabla^a\!\!~^{(1)}T_{\emm,ab}=\frac{1}{2\pi}\left[F_b{}^cR_{cd}A^d-A_b\nabla^c\left(R_{cd}A^d\right)\right] . \hspace{0.2cm}
\end{eqnarray}
In contrast, for the $\!\!~^{(0)}T_{\emm,ab}$, we have
\begin{eqnarray}
        \nabla^a\!\!~^{(0)}T_{\emm,ab}=-F_{ba}J^a-\frac{1}{2\pi}F_b{}^cR_{cd}A^d \;.
\end{eqnarray}
When the electric charge is conserved, by equation~(\ref{meq1}) we have $\nabla^c\left(R_{cd}A^d\right)=-2\pi\nabla_aJ^a=0$. Then, for the total $T_{\emm,ab}$ we have just the Lorentz force law:
\begin{eqnarray}
        \nabla^aT_{\emm,ab}=-F_{ba}J^a \;.
\end{eqnarray}
Note that, although the effect of $\!\!~^{(1)}T_{\emm,ab}$ cannot be measured by electromagnetic experiments, $\!\!~^{(1)}T_{\emm,ab}$ affects the spacetime geometry according to the Einstein field equation. Hence, $\!\!~^{(1)}T_{\emm,ab}$ represents a kind of dark electromagnetic energy and momentum \cite{li15b}.

\section{Boundary Conditions}

To have a well-defined electromagnetic field on the brane, $A^a$ and $\nabla_aA_b$ must be continuous across the brane. That is, we must have $[A^a]=0$, $[F_{ab}]=[H_{ab}]=[\Psi_{ab}]=0$. Then, by equation (\ref{Kab_H}), we have $[K_{ab}]=[\dot{g}_{ab}]$, and $[\Phi_{ab}]=-g_{ab}{}{}^{cd}[\dot{g}_{cd}]=-g_{ab}{}{}^{cd}[K_{cd}]$. By equation (\ref{join1}), we then get 
\begin{eqnarray}
        [\Phi_{ab}]=\t{\kappa}S_{ab} \;. \label{joinx}
\end{eqnarray}
In the theory of brane world, it is usually assumed that $K_{ab}^+=-K_{ab}^-$, i.e., $K_{ab}$ is antisymmetric about the brane \cite{shi00,maa10}. This $Z_2$-symmetry does not apply when the electromagnetic field is present, by equation (\ref{Kab_H}) and the condition $[H_{ab}]=0$. However, since $[K_{ab}]=[\dot{g}_{ab}]$, we can assume that 
\begin{eqnarray}
        \dot{g}_{ab}^+=-\dot{g}_{ab}^- \;,
\end{eqnarray}
i.e., $\dot{g}_{ab}$ is antisymmetric about the brane. This condition is equivalent to $\Phi_{ab}^+=-\Phi_{ab}^-$. With this assumption, from equation (\ref{joinx}) we get the boundary condition
\begin{eqnarray}
        \Phi_{ab}^+=-\Phi_{ab}^-=\frac{1}{2}[\Phi_{ab}]=\frac{\t{\kappa}}{2}S_{ab} \;. \label{join2}
\end{eqnarray}

By the symmetry properties of $\Psi_{ab}$ and $\Phi_{ab}$, we have $\dot{\Psi}_{ab}^+=-\dot{\Psi}_{ab}^-$, and $\dot{\Phi}_{ab}^+=\dot{\Phi}_{ab}^-$. Then, by equations (\ref{Tab_em_M})--(\ref{Tab_int_M}), we have $[T_{\emm,ab}]=[T_{m,ab}]=0$, and
\begin{eqnarray}
	\kappa T_{\intt,ab}^+ = -\kappa T_{\intt,ab}^-=-2\dot{\Psi}_{ab}^+-\t{\kappa}\nabla^c\left(2A_{(a}S_{b)c}-A_cS_{ab}\right) +\t{\kappa}\left[\frac{1}{6}S\Psi_{ab}-\frac{1}{3}\Psi S_{ab}-\frac{1}{2}\left(S_{cd}\Psi^{cd}-\frac{1}{3}S\Psi\right)g_{ab}\right] \;, \label{T_int_join}
\end{eqnarray}
where we have omitted the indexes ``$+$'' and ``$-$'' for $\Psi_{ab}$ and $A_a$ since $\Psi_{ab}^+=\Psi_{ab}^-$ and $A_a^+=A_a^-$. To have a well-defined Einstein field equation on the brane, each term on the right-hand side of equation (\ref{Tab}) must be symmetric about the brane, since $G_{ab}^+=G_{ab}^-$. So, we must have $\left[g_a{}^cg_b{}^d\t{T}_{cd}\right]=0$ and $\left[T_{\intt,ab}\right]=0$. Then, by equation (\ref{T_int_join}), we must have $T_{\intt,ab}^+ = T_{\intt,ab}^-=0$, i.e.,
\begin{eqnarray}
         \dot{\Psi}_{ab}^+=\t{\kappa}\left[-\frac{1}{2}\nabla^c\left(2A_{(a}S_{b)c}-A_cS_{ab}\right) +\frac{1}{12}S\Psi_{ab}-\frac{1}{6}\Psi S_{ab}-\frac{1}{4}\left(S_{cd}\Psi^{cd}-\frac{1}{3}S\Psi\right)g_{ab}\right]\;. \label{T_int_join2}
\end{eqnarray}

Let us consider a simple case: $S_{ab}=-\lambda g_{ab}$, where $\lambda$ is constant on the brane but can be a function of $w^\prime$, and $\lambda(-w^\prime)=\lambda(w^\prime)$. That is, the brane has a positive tension (represented by $\lambda$) and is vacuum otherwise. By equation (\ref{T_int_join}) and the above discussions, we get $\kappa T_{\intt,ab}^+ =-2\dot{\Psi}_{ab}^++(\t{\kappa}/3)\lambda\Psi_{ab}=0$, so we must have $\dot{\Psi}^+_{ab}=-\dot{\Psi}^-_{ab}=(\t{\kappa}/6)\lambda\Psi_{ab}$. Substituting $S_{ab}=-\lambda g_{ab}$ into equation (\ref{Tab_m_M}), we get $\kappa T_{m,ab}=-2\dot\Phi_{ab}$. If we assume that the relation in equation (\ref{join2}) holds in a small neighborhood of the brane, we get $2\dot{\Phi}_{ab}=-\t{\kappa}\left(\dot{\lambda}g_{ab}+\lambda\dot{g}_{ab}\right)=-\t{\kappa}\left(\dot{\lambda}-\t{\kappa}\lambda^2/6\right)g_{ab}$, and 
\begin{eqnarray}
        \kappa T_{m,ab}=\t{\kappa}\left(\dot{\lambda}-\frac{1}{6}\t{\kappa}\lambda^2\right)g_{ab} \;,
\end{eqnarray}
corresponding to a cosmological constant term. Then, we get the four-dimensional gravitational field equation on the brane
\begin{eqnarray}
        G_{ab}+\Lambda_\eff\, g_{ab}=\kappa T_{\emm,ab} \;, \label{ein_eq4a}
\end{eqnarray}
where we have assumed $\t{\cal T}_{ab}=0$ (eq.~\ref{tT_tSab}) and
\begin{eqnarray}
        \Lambda_\eff\equiv\t{\Lambda}-\t{\kappa}\left(\dot{\lambda}-\frac{\t{\kappa}}{6}\lambda^2\right) \;. \label{lam_eff}
\end{eqnarray}
It is just the four-dimensional Einstein field equation with a cosmological constant and the stress-energy of electromagnetic fields as the source.

To cancel the five-dimensional cosmological constant $\t{\Lambda}$ and have $\Lambda_\eff=0$, we must have $\t{\kappa}\left(\dot{\lambda}-\t{\kappa}\lambda^2/6\right)=\t{\Lambda}$. This result agrees with that discussed in \cite{shi00} when $\dot{\lambda}=0$. If on the brane $\dot{\lambda}\neq 0$ but $\t{\kappa}^2\lambda^2=-6\t{\Lambda}$, we get a residual cosmological constant $\Lambda_\eff=-\t{\kappa}\dot{\lambda}$.

To find out the relation between the coupling constants $\kappa$ and $\t{\kappa}$, let us include a stress-energy tensor of normal matter in $S_{ab}$ and denote it by $\tau_{ab}$: $S_{ab}=-\lambda g_{ab}+\tau_{ab}$. Then we get
\begin{eqnarray}
        \kappa T_{m,ab}=\t{\kappa}\left(\dot{\lambda}-\frac{\t{\kappa}}{6}\lambda^2\right)g_{ab}+\frac{\t{\kappa}^2}{6}\lambda\tau_{ab}-\t{\kappa}\dot{\tau}_{ab}+\t{\kappa}^2\pi_{ab} ,\hspace{0cm} \nonumber\\ \label{T_m_ab}
\end{eqnarray}
where
\begin{eqnarray}
        \pi_{ab}=-\frac{1}{2}\tau_{ac}\tau_b{}^c+\frac{1}{12}\tau\tau_{ab}-\frac{1}{8}\left(\tau_{cd}\tau^{cd}-\frac{1}{3}\tau^2\right)g_{ab} \hspace{0.4cm}\label{pi_ab}
\end{eqnarray}
contains quadratic terms of $\tau_{ab}$. The requirement that the linear term of $\tau_{ab}$ in (\ref{T_m_ab}) is  identical to that contained in the standard four-dimensional Einstein field equation leads to 
\begin{eqnarray}
        \kappa=\frac{1}{6}\t{\kappa}^2\lambda \;,
\end{eqnarray}
in agreement with the result in \cite{shi00}. Hence, when $\Lambda_\eff=0$, we get the general four-dimensional gravitational field equation on the brane
\begin{eqnarray}
        G_{ab}=\kappa \left(T_{\emm,ab}+\tau_{ab}\right)+\t{\kappa}\left({\cal T}_{ab}-\dot{\tau}_{ab}\right)+\t{\kappa}^2\pi_{ab} \;, \hspace{0.2cm}\label{ein_eq4}
\end{eqnarray}
where ${\cal T}_{ab}=g_a{}^cg_b{}^d\t{\cal T}_{cd}$ is the projection of the bulk stress-energy tensor.

The difference between the $\pi_{ab}$ in equation (\ref{pi_ab}) and the $\pi_{\mu\nu}$ in \cite{shi00} is caused by the fact that in our treatment the tensor $E_{ab}$ in \cite{shi00} has been expressed in terms of $K_{ab}$ and its derivative. In fact, the $\dot{\tau}_{ab}$ term in equation (\ref{ein_eq4}) arises from the expression for $E_{ab}$. (Details for the relation between $E_{ab}$ and $K_{ab}$ can be found in \cite{li15a}.)

In equation (\ref{ein_eq4}), the term linear in $\tau_{ab}$ is the stress-energy tensor of normal matter (other than electromagnetic fields) on the brane, which has the same form as in the standard Einstein field equation. It guarantees that the Newtonian gravitational law can be obtained in linear perturbations. The $\pi_{ab}$, which contains quadratic terms of $\tau_{ab}$, is important only in high energy states \cite{shi00,maa10}. The $T_{\emm,ab}$, defined by equation (\ref{Tab_em_M}), represents the stress-energy tensor of electromagnetic fields, which differs from the standard electromagnetic stress-energy tensor by a term $\!\!~^{(1)} T_{\emm,ab}$ (eq.~\ref{T_em_1}). The $\dot{\tau}_{ab}$ is the gradient of $\tau_{ab}$ with respect to the extra dimension $w$, which should be small since its effect has never been detected in normal experiments of gravity. The ${\cal T}_{ab}$ arises from the stress-energy tensor of matter in the bulk space, whose effect on the brane may look like some kind of dark matter.

\section{Relation to Other Work in the Literature}

To understand the boundary condition better, we express $K_{ab}$ in terms of $\Phi_{ab}$ and $\Psi_{ab}$
\begin{eqnarray}
        K_{ab}=-\hat{\Phi}_{ab}-\hat{\Psi}_{ab} \;, \label{Kab_Phi_Psi}
\end{eqnarray}
where
\begin{eqnarray}
        \hat{\Phi}_{ab}\equiv\Phi_{ab}-\frac{1}{3}\Phi g_{ab} \;, \hspace{1cm}
        \hat{\Psi}_{ab}\equiv\Psi_{ab}-\frac{1}{3}\Psi g_{ab} \;. \label{hat_Phi_Psi}
\end{eqnarray}
In our model, we choose the boundary condition so that $\hat{\Psi}_{ab}$ is symmetric about the brane, but $\hat{\Phi}_{ab}$ is antisymmetric about the brane. The $\hat{\Psi}_{ab}$ is interpreted as representing electromagnetic fields on the brane. The $\hat{\Phi}_{ab}$ is related to the stress-energy tensor of matter on the brane through the Israel junction relation (eq.~\ref{join2}).

The boundary condition adopted in this paper is essentially a non-$Z_2$ symmetric boundary condition, which has been studied in the framework of an asymmetric brane world in the literature (\cite{bat01,yam07}, and references therein). In \cite{bat01} and \cite{yam07}, the extrinsic curvature tensor $K_{ab}$ of the brane is separated into two parts, an antisymmetric part and a symmetric part about the brane. As usual, the antisymmetric part is related to the stress-energy tensor of matter confined in the brane by the Israel junction relation. The symmetric part is solved from a constraint equation derived from the requirement that $\left[R_{ab}\right]=0$, with a formal solution determined by the stress-energy tensor of matter confined in the brane and the antisymmetric part about the brane of the geometric quantity in the bulk space (eqs.~31 and 32 in \cite{bat01}, eq.~3.17 in \cite{yam07}). We call it ``a formal solution'' because the $\left[{\cal F}_{ab}\right]$ in \cite{bat01,yam07} contains the Weyl term $\left[{\cal E}_{ab}\right]$, which itself is a function of the $\langle K_{ab}\rangle$ and its derivative in the direction of the extra-dimension according to equation (B35) of \cite{li15a}. Electromagnetic fields are not discussed in \cite{bat01,yam07}.

Mathematically, the result in this paper agrees with that in \cite{bat01} and \cite{yam07}. The four-dimensional Einstein field equation (\ref{ein_eq4}) mathematically agrees with the four-dimensional Einstein field equation derived in \cite{bat01,yam07}. The electromagnetic field equation (\ref{meq1}) is mathematically equivalent to the Gauss-Codacci relation. This is not surprising, since in both models (the model in this paper and the model in \cite{bat01,yam07}) the effective field equations on the brane hypersurface are derived from the five-dimensional Einstein field equation by projection, and the boundary conditions on the brane are mathematically equivalent. In particular, the electromagnetic field equation (\ref{meq1}) is derived from the Gauss-Codacci relation.

However, the physics represented by our model is different from that represented by the model in \cite{bat01,yam07}. In our model, aside from the part resulted from projection of the bulk stress-energy tensor, the stress-energy tensor in the Einstein field equation on the brane contains two parts. One part, represented by $\tau_{ab}$, $\dot{\tau}_{ab}$, and $\pi_{ab}$ in equation (\ref{ein_eq4}), is interpreted as the stress-energy tensor of normal matter confined in the brane and related to the antisymmetric part of $K_{ab}$ (i.e., $-\hat{\Phi}_{ab}$) by the Israel junction relation (eqs.~\ref{Tab_m_M} and \ref{T_m_ab}). The other part, represented by $T_{\emm,ab}$ in equation (\ref{ein_eq4}), is interpreted as the stress-energy tensor of electromagnetic fields and related to the symmetric part of $K_{ab}$ (i.e., $-\Psi_{ab}$; see eq.~\ref{Tab_em_M}). The $T_{\emm,ab}$ differs from the standard electromagnetic stress-energy tensor $^{(0)}T_{\emm,ab}$ by a $^{(1)}T_{\emm,ab}$, see equations (\ref{T_em_0})--(\ref{T_em_1}). As explained in \cite{li15a,li15b}, the $^{(1)}T_{\emm,ab}$ arises from the curvature-coupled term in the electromagnetic field equation (\ref{meq1}).

Our interpretation of $T_{\emm,ab}$ as the stress-energy tensor of electromagnetic fields is motivated by the fact that the vector field equation, i.e., the Gauss-Codacci relation (\ref{vector_eq}), can be expressed in a form very similar to the Maxwell equation (eq.~\ref{meq1}). By identities
\begin{eqnarray}
        \nabla_aK^{ab}-\nabla^bK=-\nabla_a\Psi^{ab}-\nabla_a\Phi^{ab}
\end{eqnarray}
and
\begin{eqnarray}
        \nabla_a\Psi^{ab}=\nabla_aF^{ab}+2R^b{}_aA^a \;, \label{Psi_F}
\end{eqnarray}
equation (\ref{meq1}) is easily derived from equation (\ref{vector_eq}). The electromagnetic field $F_{ab}$ is related to the $\Psi_{ab}$ by equation (\ref{Psi_F}). Equation (\ref{meq1}) differs from the Maxwell equation by a curvature-coupled term, $2R^b{}_aA^a$. In a flat spacetime with a vanishing spacetime curvature, equation (\ref{meq1}) is exactly the Maxwell equation. Hence, it is natural to consider equation (\ref{meq1}) as generalization of the Maxwell equation in a curved spacetime \cite{li15a,li15b}.

The Gauss-Codacci relation was mentioned and briefly discussed in \cite{bat01,yam07}, but its relation to the Maxwell equation was not noticed. The authors of \cite{bat01,yam07} did not find the electromagnetic field contained in the field equations on the brane. Derivation of the electromagnetic field equation and identification of the electromagnetic part in the total effective stress-energy tensor in the four-dimensional Einstein field equation are the major new contribution of the present work.

\section{Summary and Discussion}

We have shown that, similar to the case in the KK theory, electromagnetic fields on a four-dimensional brane can be derived from the gravity in the five-dimensional bulk spacetime. The electromagnetic field is contained in the extrinsic curvature tensor of the brane hypersurface and obeys the field equation (\ref{meq1}), which differs from the Maxwell equation by a curvature-coupled term. Since by definition $N^a$ and $A^a$ are vectors tangent to the brane hypersurface, the electromagnetic field can only be seen on the brane and hence its effect is naturally confined on the brane. When $R_{ab}\neq 0$ the field equation is not gauge invariant. However, in a Ricci-flat spacetime with $R_{ab}=0$, the field equation (\ref{meq1}) becomes the Maxwell equation and gauge symmetry is restored. Hence, the electromagnetic field equation (\ref{meq1}) can be considered as generalization of the Maxwell equation to a curved spacetime, as an alternative to the Einstein-Maxwell equation. The curvature-coupled term $2R^b{}_aA^a$ can be regarded as a pseudo-charge current vector, whose effect is testable in an environment with high mass and energy density \cite{li15a,li15b}. 

With appropriate boundary conditions (eq.~\ref{join2}), the four-dimensional Einstein field equation is derived, which is given by equation (\ref{ein_eq4}). The right-hand-side of the Einstein field equation explicitly contains the stress-energy tensor of electromagnetic fields defined by equation (\ref{Tab_em_M}). By the relation 
\begin{eqnarray}
        \t{R}=R-K_{ab}K^{ab}+K^2-2\t{\nabla}_av^a \;,
\end{eqnarray}
where $v^a=Kn^a-a^a$, the five-dimensional Einstein-Hilbert action can be written as
\begin{eqnarray}
        S_\EH=\int \sqrt{-g}\left(L_G+L_\emm+L_m+L_\intt\right)d\t{V} \;, 
\end{eqnarray}
where $d\t{V}=2^{-1}dx^0dx^1dx^2dx^3dw^\prime$, $L_G=R$, $L_m=-g^{abcd}\dot{g}_{ab}\dot{g}_{cd}$, $L_\intt=-4g^{abcd}A_a\nabla_b\dot{g}_{cd}$, and
\begin{eqnarray}
        L_\emm = -4\sqrt{-g}\,\left(\frac{1}{4}F_{ab}F^{ab}-R_{ab}A^aA^b\right) \;.
\end{eqnarray}
It can be verified that the variation of $S_\EH$ with respect to $A_a$ leads to the electromagnetic field equation (\ref{meq1}). The variation of $S_\EH$ with respect to $g^{ab}$ leads to the four-dimensional Einstein field equation with the stress-energy tensor given by equations (\ref{Tab_em_M})--(\ref{Tab_int_M}) (see \cite{li15a} for detail).

The number of degrees of freedom (d.o.f) of gravity determined by the four-dimensional Einstein field equation is two. The number of d.o.f of the electromagnetic field determined by equation (\ref{meq1}) is three, since the presence of the curvature-coupled term causes violation of gauge symmetry. This fact means that the curvature-coupled term in the electromagnetic field equation causes an effective mass to photons. Hence, the total number of d.o.f is five, which is equal to the number of d.o.f of the five-dimensional gravity. 

So far we have not discussed the scalar constraint equation (\ref{scalar_eq}) yet. In fact, it can be replaced by another scalar equation obtained from the identity $E_a{}^a=0$, or equivalently, from substitution of equation (\ref{scalar_eq}) into the trace of equation (\ref{tensor_eq}). So, the scalar constraint equation gives essentially a constraint on the trace of $\t{\pounds}_nK_{ab}$ \cite{li15a}. It can be derived that equation (\ref{scalar_eq}) leads to
\begin{eqnarray}
        S_{ab}\Psi^{ab}-\frac{1}{3}S\Psi=-\left[\t{T}_{ab}n^an^b\right] \;,
\end{eqnarray}
where the right-hand side is the difference in the bulk pressure acting on the two sides of the brane.

Similarly, from equation (\ref{vector_eq}) (i.e., eq.~\ref{meq1}) we can derive that $\nabla_aS^{ab}=0=\nabla_a\tau^{ab}$, if $\left[g^{ab}\t{T}_{ac}n^c\right]=0$. This is just the conservation equation for $S_{ab}$ and $\tau_{ab}$. Then $\nabla^a\Phi^+_{ab}=0$, and by equation (\ref{Ja}) we have $J^a=(\t{\kappa}/4\pi)g^{ab}\t{T}_{bc}n^c$. When $J^a=0$, we have $\nabla^a{\cal T}_{ab}=0$ and $\nabla^aT_{\emm,ab}=0$, then by equation (\ref{ein_eq4}) we get $\nabla^a\dot{\tau}_{ab}=\t{\kappa}\nabla^a\pi_{ab}$.

\begin{acknowledgments}
This work was supported by the National Basic Research Program (973 Program) of China (Grant No. 2014CB845800) and the NSFC grant (No. 11373012).
\end{acknowledgments}

\end{document}